\documentclass[]{aa}  
\usepackage{graphicx}
\usepackage{xcolor}
\usepackage{amsmath}
\usepackage{epstopdf}
\usepackage{natbib}

\usepackage{txfonts}
\usepackage{hyperref}
\newcommand{\sunrise}{\textsc{Sunrise}}

\begin{document}

   \title{Cancellation of Small-Scale Magnetic Features}


   \author{A. J. Kaithakkal
          \inst{1,}\thanks{Current address: Kiepenheuer Institute for Solar Physics, Sch\"oneckstrasse 6, 79104 Freiburg, Germany; 
                                    \email{kaithakkal@leibniz-kis.de}}
          \and
          S. K. Solanki\inst{1,2}
          }

   \institute{Max Planck Institute for Solar System Research,
                 Justus-von-Liebig-Weg 3, G\"ottingen 37077, Germany
         \and
             School of Space Research, Kyung Hee University, Yongin, Gyeonggi, 446-701, Korea\\             
             }

   \date{Received ; accepted}

 
  \abstract
   {}
   {We investigate small-scale flux cancellations in a young active region observed with the high-resolution imaging magnetograph IMaX on the \sunrise{} balloon-borne solar observatory.}
   {The observed Stokes profiles of the photospheric \ion{Fe}{i} 5250.2 {\AA} line are inverted using the SPINOR code to obtain the atmospheric parameters, including magnetic field vector and the line-of-sight velocity. We then identified 11 opposite-polarity cancelling pairs using an automatic detection code, studied their evolution in detail, and derived their statistical properties. We classified the cancellations into two groups. Class I events are those for which cancellation happens between a pre-existing large magnetic feature of one polarity and a smaller feature of the other polarity that emerged/appeared nearby. For Class II events cancellations occur between two pre-existing, previously unconnected features that converge toward each other.}
   {All studied events have an apparent cancellation time less than 10 minutes and display a significant transient linear polarization signal along the polarity inversion line. The cancellation events are characterized by a flux decay rate of about 10$^{15}$~Mx s$^{-1}$. For Class I events, the Doppler velocity of the disappearing patch gradually switches from blueshift during the initial phase of cancellation to redshift towards the end of the cancellation. For class II events, the Doppler velocity is consistently redshifted. Horizontal convergence speeds of Class II pairs fall between 0.3 and 1.22~km s$^{-1}$. The elements often do not converge directly towards each other, so that the proper motion speeds of the individual elements is higher, in the range of 1 - 2.7~km s$^{-1}$.}
 {We propose that these cancellation events result from either field-line submergence (Class I), or reconnection followed by submergence (Class II and/or Class I). Ohmic dissipation of magnetic energy could also play a role for both classes. The dynamics and evolution of these events are influenced by neighbouring granular motions. We propose that, at least for the Class II events, the granular motions could possibly be driving magnetic reconnection, rather than the supergranular motions proposed for the larger cancellation events studied earlier. Specific flux cancellation rates of the Class II events seem to indicate that they belong to somewhat different category of cancellations when compared with those studied in SOT/Hinode and MDI/SOHO data.}

  \maketitle


\section{Introduction}
Magnetic flux cancellation occurs frequently in the active and quiet regions of the Sun. Cancellation leads to in situ disappearance of magnetic flux from the solar photosphere as a result of the interaction between opposite-polarity magnetic elements \citep{livi, mart}. It is a key process that removes flux from the photosphere, thus maintaining the surface flux budget \citep{sh}. \cite{cat} found from a statistical analysis that magnetic elements formed via convective collapse are mostly destroyed by flux cancellation. Cancellation is also observed to play a significant role in many of the upper atmospheric dynamic phenomena such as flares, filament eruptions and coronal mass ejections \citep[e.g.,][]{wn, zh, yar}, X-ray bright points \citep{pr, har, jz}, Ellerman bombs \citep{geo, wat, vs, rd}, the formation of prominences \citep{martb}, coronal jets \citep{pn}, etc. 

If the two cancelling magnetic features were already connected prior to cancellation, then the disappearance of flux can occur without prior magnetic reconnection, e.g. by the retraction of a pre-existing $\Omega$ loop. In most cases, however, reconnection is expected to occur as cancellation proceeds. Two dynamically different mechanisms, both usually involving magnetic reconnection, are invoked to describe the observed flux cancellation events -- $\Omega$-loop submergence or U-loop emergence across the solar surface \citep{zw}. If reconnection occurs above the solar surface, the canceling features disappear from the surface accompanied by a descending $\Omega$-loop; if two opposite polarity magnetic features come together and reconnection happens below the surface, a U-loop emerging subsequently through the solar surface results in the removal of opposite polarity elements \citep[cf.][]{spr, mag}. 

In both the cases, horizontal magnetic field is expected to be present between the opposite-polarity features as cancellation proceeds. Information on the vector magnetic field and on the vertical motion of the transverse magnetic field is used to distinguish between the two scenarios. The horizontal fields move downward in $\Omega$-loop submergence, and upward in U-loop emergence. Using simultaneous photospheric and chromospheric observations \cite{harb} found that about half (44\%) of the cancellation events they investigated support $\Omega$-loop submergence, while a minor fraction (18\%) represents emergence, and the rest (38\%) could not be classified owing to the low cadence and noise. Observations of a horizontal magnetic field and redshift at the cancellation sites that support flux submergence were reported in a number of other studies as well \citep[eg.,][]{cha, yn, zha, iida}. Cancellation events interpreted as the emergence of a U-loop are described in the investigations by, e.g., \cite{van}, \cite{yu}, \cite{bel}. Ellerman bombs, which are thought to be triggered by flux cancellation \citep[e.g.,][]{hash, rd}, also involve U-shaped loops, with reconnection happening within the U-loop.

Granular-scale flux cancellation events studied by \cite{kbb} using the Solar Optical Telescope \citep[SOT;][]{ts} aboard Hinode \citep{ks} were found to proceed in general without the presence of transverse magnetic field between the colliding magnetic elements. The authors attribute the absence of horizontal magnetic field to the limited spatial resolution of the SOT/SP. They proposed that the Doppler velocity at these cancellation sites does not necessarily represent the motion of the field lines. Rather, they characterize the convective motion in the region where the cancellation takes place. However, another study \citep{catb} using Hinode/SP data, which has a cadence of ~1 minute, found that about 80\% of the cancellation events they investigated are associated with transverse magnetic field.

Small-scale, short-lived magnetic features that carry a substantial amount of magnetic flux are observed almost everywhere on the solar surface. Their emergence, evolution and eventual disappearance from the surface are of importance for obtaining insight into the causes of solar activity. However, detailed investigation of small-scale dynamics is limited by spatial-resolution constraints. This is exactly where our study becomes significant. During its second flight, the balloon-borne solar observatory \sunrise{} \citep{bart,gan,berk,samia,samib} acquired high spatial resolution and high cadence observations of an emerging active region. In the full field of view (FOV) of the Imaging Magnetograph eXperiment \citep[IMaX,][]{pillet} instrument onboard the \sunrise{} balloon-borne solar observatory, we could identify a number of sub-regions where cancellation was happening. We studied these small-scale cancellation events and derived their statistical properties, as described in the following sections.

\section{Observations and Data Analysis} \label{sec:obsv}
\begin{figure*}
\centering
  \includegraphics[trim=3 335 140 194,clip,width=0.8\textwidth]{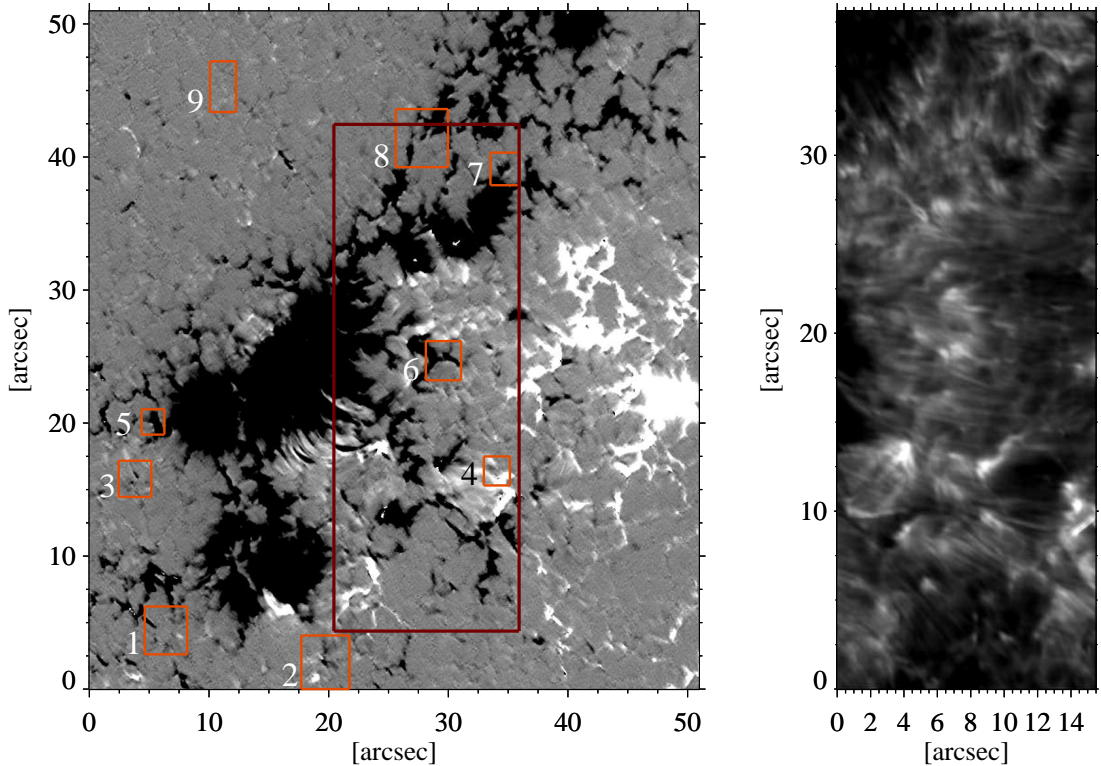}
  \caption{Left: B$_\mathrm{LOS}$ map saturated at $\pm$ 350~G obtained by IMaX on \sunrise{}. The maroon rectangle represents the SuFI FOV and the red boxes outline our regions of interest (ROI, numbered from 1 to 9). Right: Intensity image from the SuFI channel at \ion{Ca}{ii} {H} 396.8~nm.}
  \label{fig:a}
\end{figure*}

The observations were carried out on 12 June 2013  between 23:39:10 and 23:55:37 UT with a cadence of 36.5~s. IMaX recorded polarized spectra of the photospheric line \ion{Fe}{i} 5250.2 {\AA} (Land\'{e} factor, g = 3) at eight wavelength points -- $\lambda$ = $\pm$120, $\pm$80, $\pm$40, 0, and 227 m{\AA} from the line center -- with four accumulations at each position. The FOV of 51$\arcsec$ x 51$\arcsec$ covered a young active region (see, Fig. \ref{fig:a}) AR 11768 (located at $\mu$ = 0.93) with a scale of 0.0545\arcsec per pixel. 

At the same time, the Sunrise Ultraviolet Filter Imager \citep[SuFI,][]{gan} collected intensity images of the photosphere and lower chromosphere in a smaller FOV (maroon rectangle in the left panel of Fig. \ref{fig:a}) with a pixel sampling of 0.02\arcsec. SuFI recorded intensity images at three wavelengths (a broadband channel in the UV continuum around 300~nm and two narrower bands in the core of the \ion{Ca}{ii} {H} 396.8~nm line) at a cadence of about 7~s for the set of three images. The right panel in Fig. \ref{fig:a} presents the \ion{Ca}{ii} {H} intensity image at 396.8~nm at the beginning of the observation.

The IMaX data were reconstructed with the help of a point spread function retrieved from inflight phase-diversity measurements \citep{pillet}. The reconstructed data have a spatial resolution of 0.15\arcsec -- 0.18\arcsec. The noise ($\sigma$) in Stokes $V$ is about 7$\times$10$^{-3}I_\mathrm{C}$, where $I_\mathrm{C}$ is the continuum intensity. The vector magnetic field and line-of-sight (LOS) velocity ($v_\mathrm{LOS}$) information are derived from inversions of the Stokes vectors using the SPINOR code \citep{fru}.  Height independent magnetic field parameters ($B$, $\gamma$ and $\phi$), the line-of-sight velocity, $v_\mathrm{LOS}$, and temperature at three heights were returned by the inversion. The inversion assumed a single atmospheric component for each pixel, so that all features are assumed to be resolved. In this sense, parameters such as the magnetic field strength are expected to be lower limits. However, due to the low signal-to-noise ratio in Stokes $Q$ and $U$, it is likely that the inclination angle is overestimated, i.e. the magnetic features are likely less transverse than they appear to be from the inversions, in particular where $Q$ and $U$ lie less than 3$\sigma$ above the noise \citep[e.g.,][]{borb}. 

More details on the data reduction, inversions, and LOS velocity calibration are provided by \cite{samib}. The signature of $p$-modes was removed from the continuum and $v_\mathrm{LOS}$ maps using a subsonic filter \citep{title} with a cut-off phase velocity of 4 km s$^{-1}$. The continuum intensity maps are normalized with respect to the mean quiet-Sun value. Image sequences of all the relevant parameters were corrected for rotation of the FOV caused by the alt-azimuth mounting of the \sunrise{} telescope and aligned using a spatial cross-correlation technique. We obtained the total linear polarization (LP) maps as follows: (1) $3\times 3$ pixel binned Stokes $Q$ and $U$ maps were retrieved from the non-reconstructed dataset, (2) these maps are then used to get the total LP signal integrated over six wavelength points within the spectral line: $\lambda$ = $\pm$120, $\pm$80 and $\pm$40~m{\AA}. Non-reconstructed data were used to obtain LP maps, as their noise level is 2.5 - 3 times lower than that of the reconstructed data. 

\subsection{Selection of Opposite Polarity Pairs} \label{sec:pairs}
We focus on the regions within the red boxes in the left panel of Fig. \ref{fig:a}. In the following the boxed areas are termed regions of interest (ROI) and they are identified in the following by the numbers written next to the boxes. A modified version of the multilevel tracking (MLT) algorithm of \cite{bov} is applied to maps of the LOS component of the magnetic field vector, $B_\mathrm{LOS}$, to automatically detect magnetic elements in these sub-regions. MLT uses threshold levels in decreasing order. We chose 20 thresholds varying from 130~G to 40~G for the MLT algorithm (the lowest threshold is 3 times the standard deviation of the quiet-Sun $B_\mathrm{LOS}$ value, obtained by considering signals only from within granules, i.e. ignoring the intergranular lanes where magnetic fields tend to concentrate). If one of the threshold levels splits a feature (which was identified by the preceding threshold as a single entity) into two, and if the separation between those two features is less than three pixels, the code combines them into one to avoid artificially splitting a feature into sub-features. Only those features that have a minimum area of five pixels were selected for further analysis (smaller features are below the spatial resolution limit). From the selected individual features, the code then identified (cancelling) opposite polarity pairs based on the condition that their boundaries are separated by at most three pixel. That is, pairs are identified and followed and studied only during the phase in which their members are very close to each other.

The code then follows each of the pairs in time. A given pair is assumed to be the same in consecutive images if there is a spatial overlap between features in consecutive images \citep[cf.][]{an}. We chose only those pairs of opposite polarity which show a decrease in magnetic flux over time and both of whose members are visible for at least 3 IMaX frames. For the purposes of the present study, the pair ceases to exist if either of the elements disappears or moves away from the other, or if it splits into two or more fragments and the largest of these carries less than 30\% of the flux of the original element. Similarly, if a member of a pair is formed by merging, then the largest merging fragment is taken as the element prior to the merging.

From the nine ROI's, we selected 11 feature pairs. These pairs were then grouped into two classes -- Class I: Pairs for which cancellation takes place between a large feature of one polarity and a smaller feature of the opposite polarity. We selected 6 pairs in this class. The total magnetic flux of the smaller feature increases and reaches a maximum during the initial phase, and then steadily decreases. Of the 6 cancellations, four happen between a newly appearing sub-arcsec feature and a pre-existing, comparatively large magnetic feature. In the remaining two pairs, both features are present from the beginning of the observation. Since these two pairs display a similar subsequent evolution as the other four, we grouped them together. For these 6 pairs, we determined the relevant physical parameters of the smaller feature as well as along the polarity inversion line (PIL) separating them. Note that the width of the PIL is one pixel.

Class II: Pairs for which cancellation happens between pre-existing, previously unrelated features which converge toward each other and cancel. 5 pairs were selected in this class. For Class II, the physical parameters along the PIL, and of the magnetic element of the pair, from which the flux decay rate is determined, were calculated. The convergence speed of the Class II pairs is determined as follows. (1) The spatial center of gravity (COG) of both the opposite polarity elements were calculated. The COG is defined as mean position of the feature weighted by magnetic flux. (2) for each frame, the separation (r) between the COG's of the pair's elements was obtained. (3) Finally convergence speed ($v_\mathrm{conv}$) was derived by performing a linear fit to the time -- separation plot, $v_\mathrm{conv} = -0.5 \times dr/dt$ \citep{chb}. We also estimated specific flux cancellation rate, $r \equiv  R /  l$, where $R \equiv d\phi/ dt$ is the flux decay rate, and $l$ is the horizontal length of the flux cancellation interface \citep[see][for details]{chb}.
\section{Results} \label{sec:results}
\subsection{Class I Pairs - case study} \label{sec:ca}
\begin{figure*}
\centering
  \includegraphics[trim=3 380 118 20,clip,width=0.8\textwidth]{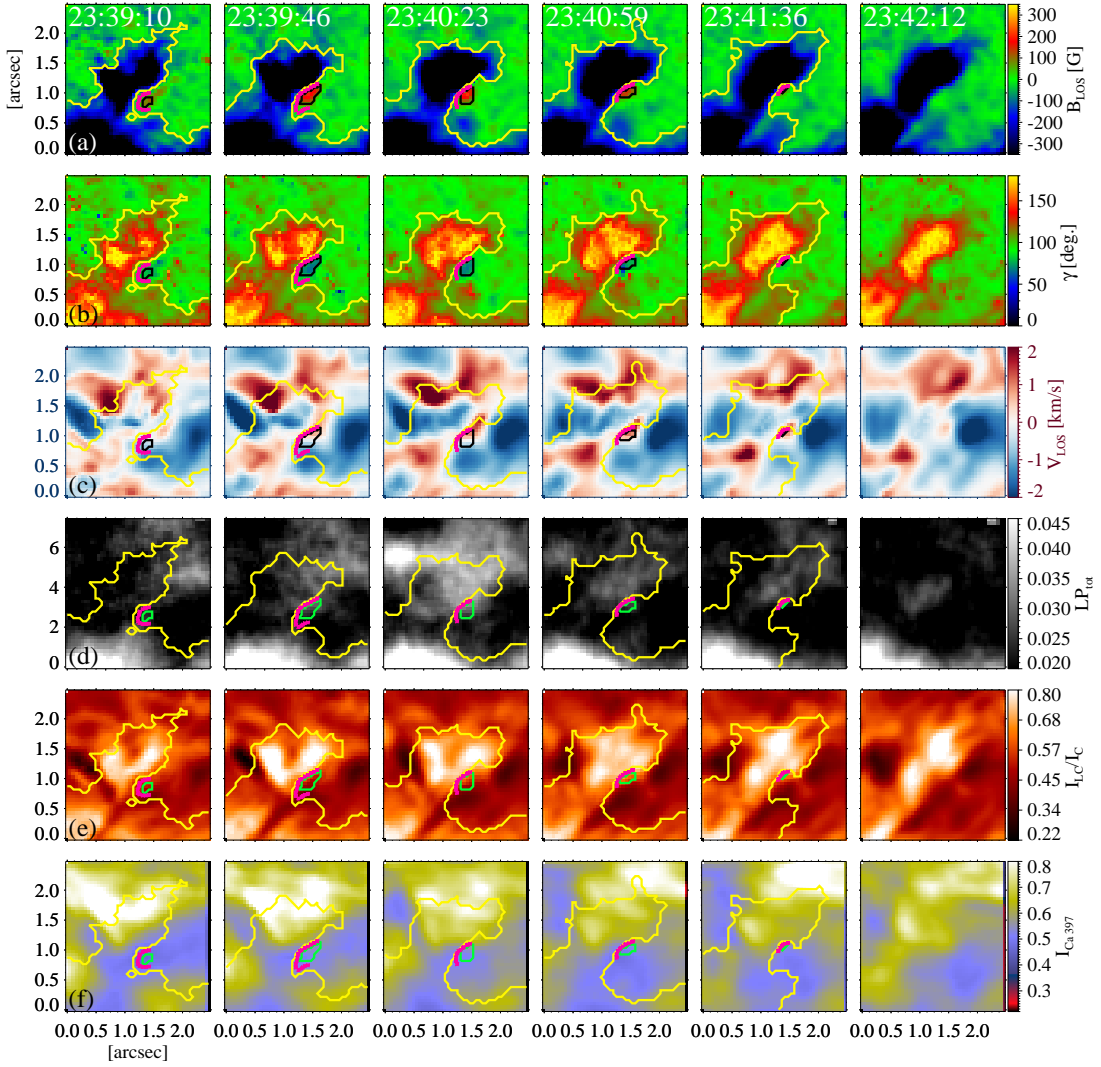}
  \caption{Evolution of the pair of magnetic elements in ROI 7. From top to bottom: Time series of (a) line-of-sight component (LOS) of magnetic field, $B_\mathrm{LOS}$, (b) field inclination relative to the LOS, $\gamma$, (c) line-of-sight velocity, $v_\mathrm{LOS}$, (d) total linear polarization ($LP_\mathrm{tot}$), (e) line core intensity $I_\mathrm{LC}$ normalized to the local continuum $I_\mathrm{C}$, and (f) \ion{Ca}{ii} {H} 397 nm intensity recorded by SuFI. The yellow contour encloses the large negative magnetic polarity patch and the black ((a) -- (c))/green ((d) -- (f)) contour indicates the small positive polarity patch of the pair. The magenta dots outline the PIL between the two patches. Times at which the images were recorded are given at the top of the panels in row (a).}
  \label{fig:b}
\end{figure*}

Before considering common features of the cancellation events, we first consider one cancelling pair in detail, viz. the pair in ROI 7 in Fig. \ref{fig:a}. The temporal evolution of various physical parameters of this pair is presented in Fig. \ref{fig:b}. Contours are plotted over features only when they are identified by the code as belonging to a cancelling pair. The $B_\mathrm{LOS}$ (Fig. \ref{fig:b} a) maps show the evolution of a small positive feature adjacent to a large negative one. This pair is present from the time the observation sequence starts, although at the beginning the flux of the smaller feature still increases. After the second frame, however, the flux constantly decreases, so that the positive patch disappears completely within about 2.4 minutes (five frames). The magnetic vector of the positive feature was found to be inclined (Fig. \ref{fig:b} b) by about 75$^\circ$ averaged over the lifetime. From the $v_\mathrm{LOS}$ maps (Fig. \ref{fig:b} c) it is clear that the positive polarity patch switches from upflow to downflow as cancellation proceeds. 

A signal is clearly seen along the PIL in the first four frames of the total linear polarization maps (Fig. \ref{fig:b} d), although not generally restricted to the PIL. During the event, the line core intensity rises along the PIL, as is visible from Fig. \ref{fig:b} e. This could either be contributed by the bright part of the large negative patch neighboring the positive one, or could be the result of reconnection between the opposite polarity elements. Even with the high-resolution IMaX data, we are unable to distinguish between these two scenarios. The \ion{Ca}{ii} {H} 397~nm intensity images from SuFI do not show any intensity enhancement while cancellation happens, which if present would have supported the second possibility (i.e., reconnection).

\begin{figure}
\centering
  \includegraphics[trim=218 278 135 240, clip, width=0.52\textwidth]{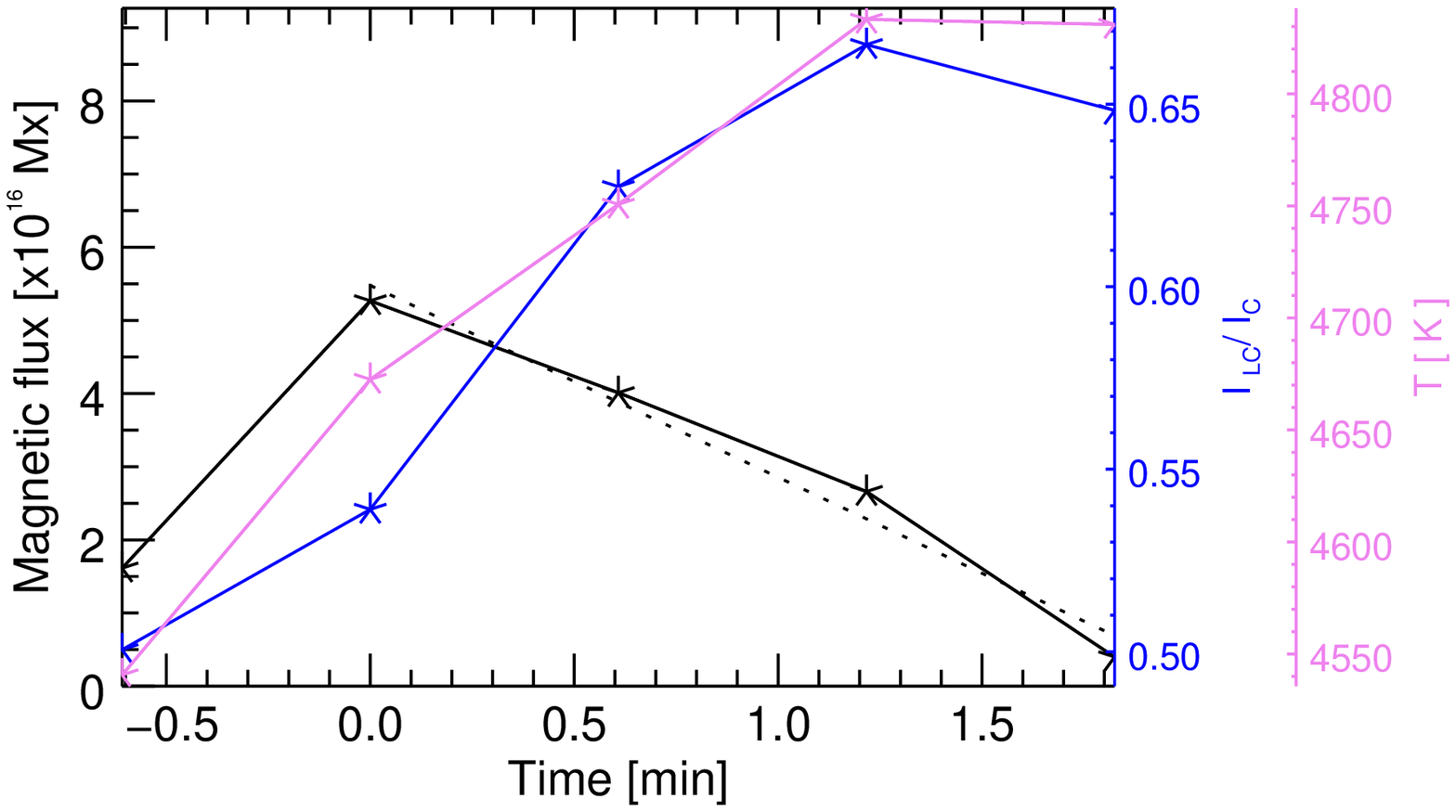}
  \caption{Evolution of key parameters of the cancelling flux in ROI 7. Plotted are the variation with time of total magnetic flux of the positive polarity patch (black, referring to the scale on the left), line core intensity (blue, inner scale on the right) and temperature at log $\tau =  \mathrm{-2.5}$ (magenta; outer scale on the right) averaged over the PIL. The line core intensity, $I_\mathrm{LC}$, is normalized to the local continuum intensity, $I_\mathrm{C}$. The dashed line is the linear fit to the decay phase of the flux evolution. Zero on the time axis represents the time at which the total magnetic flux of the smaller magnetic feature reaches its peak.}
  \label{fig:c}
\end{figure}

The black curve in Fig. \ref{fig:c} shows the evolution of total magnetic flux of the positive polarity patch. From a linear fit to the decreasing part of the flux curve we estimated the rate of flux decay to be $\sim -4.0 \times 10^{14}$ Mx s$^{-1}$ (i.e. half the rate of flux cancellation). It is difficult to calculate the flux decay rate of the larger negative polarity patch as it undergoes merging and splitting during the cancellation (at locations far from the cancellation), causing random increase or decrease in its total magnetic flux. The blue line in the same figure represents the variation of line core intensity averaged over the PIL. It rises by about 33\% in terms of $I_\mathrm{C}$ during the event. The temperature at log $\tau = \mathrm{-2.5}$ averaged over the PIL is plotted in magenta color. It shows an increase of about 290 K while the patch evolves and correlates very well with the evolution of the line-core intensity, as expected. 

\subsection{Class I Pairs - properties of the 6 pairs} \label{sec:caa}
In all the 6 pairs belonging to Class I, a small element of one polarity is observed to interact with a much bigger element of the other polarity, resulting in the complete removal of the smaller feature from the solar surface. For the pair in ROI 2, the flux-ratio between the smaller and the larger patch is 3.5\% when the smaller patch reaches its peak flux. We calculated the flux-ratio for only this region as the larger patch of the pair more or less stayed compact during the pair-lifetime. For all other pairs, the larger magnetic patch interacts with its surroundings leading to merging and splitting thus making it difficult to quantify its flux decay rate. The shapes of the disappearing patches and the Doppler velocity within them and along the PIL seem to be greatly influenced by the dynamics of the granules in the neighborhood. The $V_\mathrm{LOS}$ maps clearly show that the Doppler velocity within the patches or along the PIL mirrors the LOS convective motion in the immediate vicinity. 

Table \ref{tab:t1} details values of physical parameters that characterize the disappearing smaller feature in each of the 6 pairs. The values in columns 2 and 6 (peak flux and maximum area, respectively) are obtained by first calculating the total magnetic flux and area of the smaller patch over its lifetime and then by selecting the respective peak values during that period. For the 3rd and 5th column, peak values of $B_\mathrm{LOS}$ and $V_\mathrm{LOS}$ (single pixel values) of a disappearing patch are first determined for each time frame and then the maximum values during its life are chosen. For the 4th column, we first determined $\gamma_\mathrm{LOS}$ for each time frame, at the pixel where $B_\mathrm{LOS}$ is maximum, and from those values the one that corresponds to $B_\mathrm{LOS, max}$ (3rd column, Table \ref{tab:t1}) was obtained.

We found that the smaller features that disappear fully - (1) are of sub-arcsecond size, (2) have weak, inclined fields, and (3) undergo a change in Doppler velocity from blueshift to redshift in the course of cancellation. Along the PIL between the feature pairs we observe the following: (1) the presence of LP signal above 3$\sigma$ for at least some time during cancellation. The LP signal rises to a maximum followed by a decline towards the end of the event, (2) a rise in line core intensity as cancellation proceeds, and (3) no visible increase in intensity in SuFI images at both, 300~nm and 397~nm (\ion{Ca}{ii} {H} line core). 

We also found that the smaller feature belonging to one of the 6 pairs exhibits substantial signal (greater than 3 times the quiet-sun standard deviation value, which is an upper limit for the noise level at that wavelength) in the Stokes $V$ continuum ($V_\mathrm{C}$) map during the pair-lifetime. The $V_\mathrm{C}$ signal lasts for about 2.4 minutes (5 frames) within the smaller magnetic feature of this pair and reaches a peak value of about $0.025\times I_\mathrm{C}$. Strong Stokes $V$ continuum signal in the vicinity of opposite polarity features and inclined magnetic fields has been reported in the studies of \cite{bora, borc}, \cite{mar}, and \cite{qn}. These studies attribute $V_\mathrm{C}$ to supersonic flows accelerated by magnetic reconnection of the concerned opposite polarity features.
 
\begin{table*}
\centering
\caption{Properties of Class I Pairs}
\label{tab:t1}
\begin{tabular}{c c c c c c c c c c}  
\hline\hline    \\
ROI & Peak Flux & $B_\mathrm{LOS, max}$ & $\gamma_\mathrm{LOS}$\tablefootmark{a} & $V_\mathrm{LOS, max}$  & Area$_\mathrm{max}$ & Cancellation time\tablefootmark{b}& Flux decay rate & T\tablefootmark{c} & SuFI\tablefootmark{d}\\
   &  [10$^{16}$~Mx] & [G] & [deg.] & [km s$^{-1}$] & [arcsec$^2$] & [min] & [10$^{14}$ Mx s$^{-1}$] & K & \\\\
   \hline    \\
2\tablefootmark{e}  & 4.3   & 153 & 44 & $-$0.66 & 0.1 & 1.2 & $-$4.0 & 230 & No\\
4 & $-$12.2 & 200 & 72 & 1.5 & 0.23 & 3.0 & $-$4.7  & 60 & Yes \\
5  & 13.9  & 290 & 65 & 2.56  & 0.18  & 3.6 & $-$5.0 & 70 & No\\ 
6 & 12.2 & 190 & 69 & 1.4 & 0.27 & 3.6  & $-$4.0 & 110 & Yes \\          
7 & 5.3 & 210 & 67 & 2.0 & 0.08 & 1.8 & $-$4.0 & 80 & Yes\\
8 & 15.4 & 215  & 67 & 1.5 & 0.39 & 6.7 & $-$2.4 & 80 & Yes\\
\hline    
\end{tabular}
\tablefoot{Properties of only the smaller features that disappear during cancellation are given.\\
\tablefoottext{a}{For a negative polarity patch $\gamma_\mathrm{LOS}$ = 180 - $\gamma_\mathrm{LOS}$.}\\
\tablefoottext{b}{Cancellation time is calculated as the difference between the time when the smaller feature reaches its peak flux and the time when it is last detected.}\\
\tablefoottext{c}{Gives enhanced temperature value along the PIL at log $\tau = \mathrm{-2.5}$}\\
\tablefoottext{d}{Indicates whether the pair is covered by the SuFI FOV.}\\
\tablefoottext{e}{Two pairs are identified in ROI 2; one belongs to Class I and the other belongs to Class II}\\
}
\end{table*}
  
\begin{figure}
\centering
  \includegraphics[trim=218 274 125 237, clip, width=0.52\textwidth]{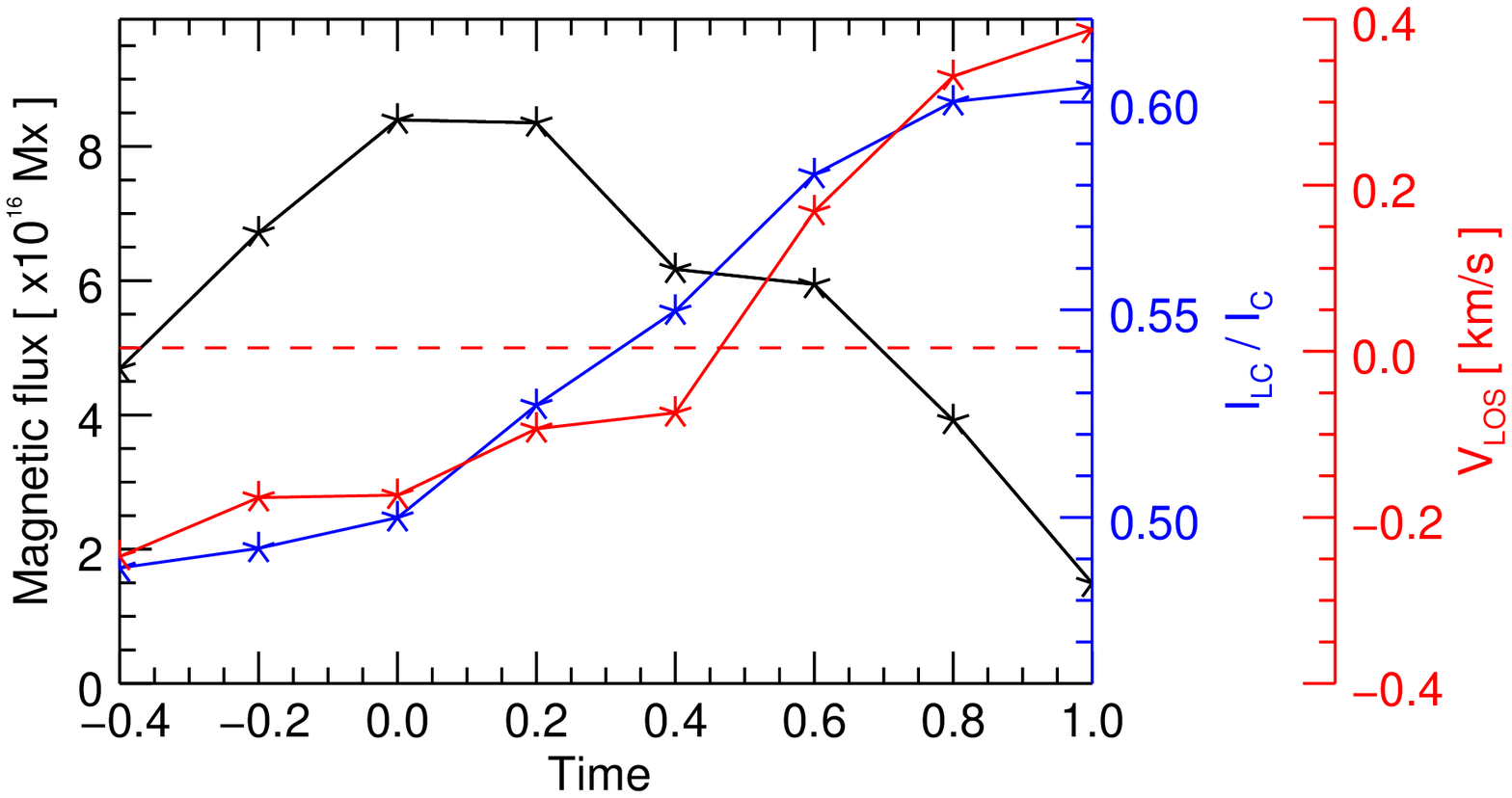}
  \caption{Averaged evolution of physical properties of the smaller features belonging to Class I pairs from birth (negative time values) to death (normalized time 1; see main text for details). The averaged flux of the smaller magnetic feature is plotted in black (scale to the left). The average Doppler velocity of the smaller magnetic features is in red (scale to the right). The dashed red line represents zero on the velocity axis. Positive values of Doppler velocity represent downflows. The blue line depicts the mean line core intensity along the PIL (scale to the right).}
\label{fig:d}
\end{figure}

The average evolution of the magnetic flux, the line of sight velocity and the line core intensity of the 6 cancelling pairs is plotted in Fig. \ref{fig:d} vs. time normalized such that the the lifetime of cancellation in each example is exactly unity. In order to be able to average together the properties taken from different Class I pairs, the pair-lifetimes are normalized, so that 0 marks the time when the smaller feature attains its peak flux value (see the x-axis of Fig. \ref{fig:c}) and 1 denotes the time step just before the smaller feature disappear. The magnetic flux of the smaller feature reaches a peak before starting to decrease during cancellation. The mean Doppler velocity of the smaller patches (red) shows a transition from blue-shift during the phase of increasing flux to red-shift as magnetic flux starts to fall off. The line core intensity along the PIL (blue) increases during the whole time, while the flux is grown and also while it is decreasing, although the increase slows down towards the end (in the example shown in Fig. \ref{fig:c}, it actually reverses). 

\subsection{Class II Pairs - case study} \label{sec:cb}
\begin{figure*}
\centering
  \includegraphics[trim=3 575 125 20,clip,width=0.8\textwidth]{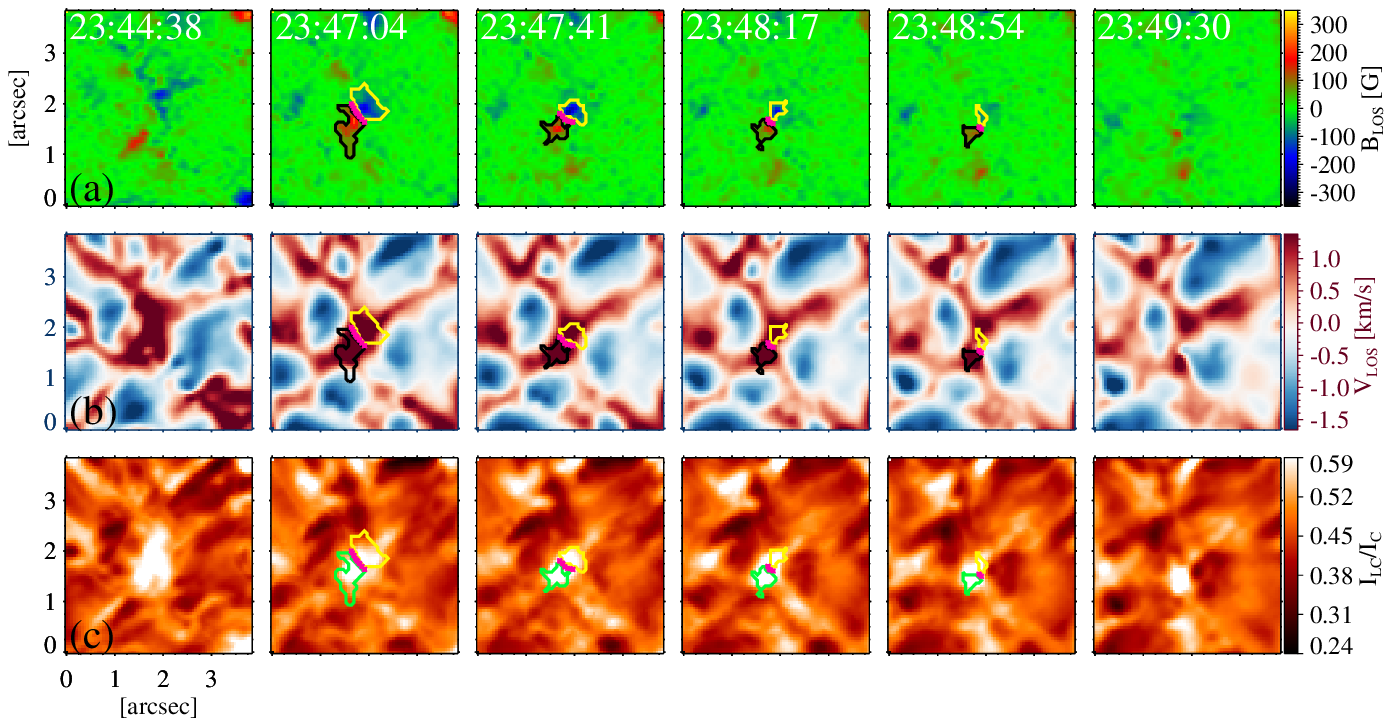}
  \caption{From top to bottom: Time series of (a) line-of-sight component of magnetic field, $B_\mathrm{LOS}$, (b) line-of-sight velocity, $v_\mathrm{LOS}$, and (c) line core intensity normalized to the local continuum, for the cancelling pair in ROI 9. The magenta dots outline the PIL. The axes scales are the same as in Fig. \ref{fig:b}. (A movie of the extended time series is available in the online journa l).}
\label{fig:e}
\end{figure*}

According to Fig. \ref{fig:a}, the Class II pairs are found mainly at the edge of the active region (pairs 1, 2, 3), or even completely outside it (pair 9). This is in contrast to the Class I pairs, which are mainly present in the active region itself.  The life history of one of the Class II pairs (occurring in ROI 9 in Fig. \ref{fig:a}) is shown in Fig. \ref{fig:e}, starting just before the pair was identified by the code, i.e. while the features of the cancelling pair are still separated. Thus $B_\mathrm{LOS}$ map at 23:44 UT (first frame of Fig. \ref{fig:e} a) shows two spatially separated features of opposite polarity. These features undergo multiple splittings and mergings (visible in the movie). At 23:47:04 UT they converge toward each other and are identified by the code as a pair. The total magnetic flux of both the features decreases rapidly during the time they touch each other. After they have lost most of their flux (74\% of that of the positive polarity feature and 86\% of the negative polarity feature) they start to drift apart again, so that at 23:49 UT they are no longer considered to be a pair by the code as the separation between them becomes more than 3 pixels. The Doppler velocity maps (Fig. \ref{fig:e} b) show that the cancellation happens consistently in a downflow lane. This is not surprising, given that most magnetic features are consistently found in downflow lanes for most of their lifetime. At 23:48:54 UT an upflow region intrudes into the location of cancellation (i.e. the PIL). The cancellation is associated with a bright patch visible in the line core intensity maps (Fig. \ref{fig:e} c). From the map at 23:44 UT, it is clear that the bright patch existed before the features started cancelling, but it became brighter in the course of the cancellation. The $B_\mathrm{LOS}$ and $I_\mathrm{LC}/ I_\mathrm{C}$ maps in the movie show that the bright patch forms as the opposite polarity features appear in the magnetogram and start approaching each other. This could indicate that the magnetic field of the features likely expands very fast with height, so that the features undergo reconnection above the lower-mid photosphere (i.e. the height at which IMaX samples the magnetic field) even before they come spatially close to each other at this height. The COG's of opposite polarity features were 1.86$\arcsec$ apart when the bright patch became clearly visible in the line core intensity map. At that instant the distance between the two closest edges of the features is 1.5$\arcsec$.

\begin{figure}
\centering
  \includegraphics[trim=219 270 117 235, clip, width=0.50\textwidth]{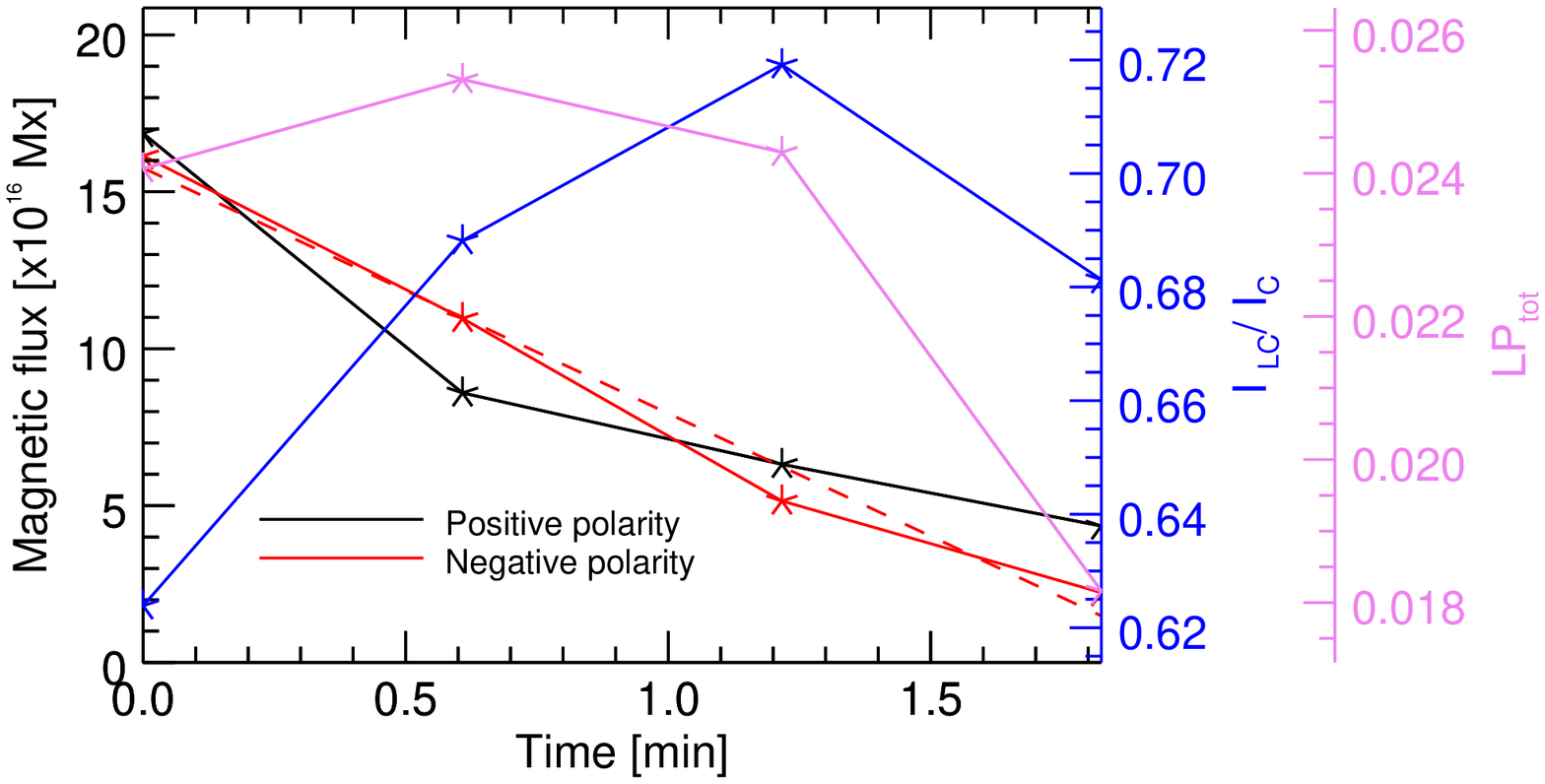}
  \caption{Same as the Fig. \ref{fig:c}, but for Class II cancellation event in ROI 9. The black (red) solid line represents the evolution of total magnetic flux of the positive (negative) patch. The dashed red line is the linear fit to the flux curve of the negative patch. Variation of total linear polarization ($LP_\mathrm{tot}$) averaged over the PIL is plotted in magenta. The normalized line-core intensity is plotted in blue.}
  \label{fig:f}
\end{figure}

The temporal variation of the total magnetic flux of the positive and negative polarity patches is plotted in Fig. \ref{fig:f}. The magnetic flux of both polarities decreases over time by a roughly similar amount, as expected for magnetic flux removal from the solar surface. From a linear fit (dashed red line) to the flux curve of the negative patch (solid red line), we obtained a flux decay rate of $-1.3 \times 10^{15}$ Mx s$^{-1}$. The normalized line core intensity (blue line) averaged over the PIL shows an increase of about 15\% before decreasing again. The decrease can be understood in terms of reduced magnetic reconnection rate as the magnetic flux patches start to move apart at the end of the plotted sequence. As the magnetic reconnection rate drops, the gas cools down. The $LP_\mathrm{tot}$ (magenta line) signal along the PIL  display values above the $3\sigma$ level during the event. We also found that the temperature at log $\tau = \mathrm{-2.5}$ averaged over the PIL rises by about 80 K during cancellation and then drops in the final frame (for the same reasons as the line core intensity) in which the pair was last identified.

\subsection{Class II Pairs - properties of the 5 pairs} \label{sec:cbb}
The characteristics of Class II pairs are detailed in Table \ref{tab:t2}. The parameters in the table are determined in the same way as described in Sec.\ref{sec:caa}. For the pair in ROI 1, a small positive polarity patch approaches a larger negative patch and disappears completely. In the case of the other 4 pairs, neither element of the pair is removed completely from the surface - after being in contact for a while, the opposite polarity elements drift away from each other again. None of the Class II pairs were covered by the SuFI FOV. Consequently, there is no column on SuFI. We chose one representative element from each pair, onto the flux evolution of which a linear fit is performed to determine the flux decay rate. The representative element is the one having least interactions with its surroundings (i.e. least affected by splitting and merging with other magnetic features). Compared to the other Class II pairs, the ones in ROI's 1 and 2 are bigger and have stronger magnetic fields. 

The Class II events are of a type that is more commonly reported in the literature \citep[e.g.,][]{chb, lit, pa, zha, kbb, iida}. The magnetic features involved in the cancellation appear to be unconnected on the solar surface in the beginning (at least we found no signs of an earlier common evolution). Then they are driven toward each other by horizontal motions that converge to an intersection of downflow lanes. The converging speed of the events falls in the range $0.3 - 1.8$~km s$^{-1}$. The convergence speed of the pair in ROI 3 is comparable to the supergranular speed, while all the other pairs converge at higher speeds. The pair in ROI 9 (see Fig. \ref{fig:a}) is the furthest from the pore, and has the highest converging speed. The other pairs that are closer to the pore converge more slowly. It could be that their motions are significantly affected by the strong magnetic field of the pore, as pointed out in \cite{title}. Some previous studies reported convergence speeds similar to typical supergranular flow speeds \citep[e.g.,][]{chb, lit, iida}. We speculate that the larger features studied by these authors are mainly affected by supergranular flows, while the smaller features we study are mainly driven by granular flows. To confirm that the individual elements of Class II pairs are indeed guided by granular motions, we calculated their proper motion speeds. We found that the proper motion speeds of the elements fall in the range of $1- 2.7$~km s$^{-1}$. Based on these values, we suggest that the Class II cancellation events are driven by granular scale motions. Values of proper motion speeds we obtained are comparable to the horizontal velocity values of internetwork bright points \citep[c.f.][and the references therein]{jaf}. There have been some earlier reports of comparable convergence speeds to those found here. \cite{park}, in their study on granule-scale canceling features, reported an event with converging speed of about $1$~km s$^{-1}$. \cite{zha} suggested that two of the six cancellations events they investigated are set off by granular flows, with the magnetic element of one polarity being advected to that of the opposite polarity with a velocity above 1~km s$^{-1}$. This explanation is consistent with our proposal, given above.

\begin{table*}
\caption{Properties of Class II Pairs}
\label{tab:t2}
\centering   
\begin{tabular}{c c c c c c c  c}
\hline\hline  \\
ROI & Peak Flux & $B_\mathrm{LOS, max}$ & $\gamma_\mathrm{LOS}$ & $V_\mathrm{LOS, max}$  & Area$_\mathrm{max}$ & Cancellation time & $v_\mathrm{conv}$\\
   & [10$^{16}$~Mx] & [G] & [deg.] & [km s$^{-1}$] & [arcsec$^2$] & [min] & [km s$^{-1}$]\\\\
   \hline    \\
1  & 39.25  &  487 & 73 & 1.7 & 0.53   & 7.3  & 1.22\\
2  & $-$107.68  & 1614 & 15 & 2.9  & 0.87  & 4.3  & 0.61\\
3  & 24.2 & 270 & 68 & 2.2 & 0.41 & 1.8  & 0.56\\ 
    & 7.3 & 187 & 66 & 1.6  &  0.20  & 3.0 & 0.30\\ 
9  & $-$16.12  & 248 & 62 & 2.0 & 0.35 &  1.8 & 1.84\\
\hline    
\end{tabular}
\end{table*}

The rates of flux decay ($R$) of the representative elements, cancellation interface length ($l$) between the pair elements, specific flux cancellation rates ($r$) of the representative elements, and ($B_\mathrm{LOS, max} \times v_\mathrm{conv}$) values for Class II are given in Table \ref{tab:t3}. The first three quantities are defined at the end of Sec.\ref{sec:pairs} and values for the last column are taken from columns 3 and 8 of Table \ref{tab:t2}. We find that ($B_\mathrm{LOS, max} \times v_\mathrm{conv}$) values are comparable to values of $r$. According to \cite{chb} and \cite{pa}, the product of $B_\mathrm{LOS, max}$ and $v_\mathrm{conv}$ is comparable to  r,  if $B_\mathrm{LOS, max}$ is close to the field strength in the inflow region and $v_\mathrm{conv}$ is close to the inflow speed. 

\begin{figure}
\centering
  \includegraphics[trim=200 273 85 238, clip, width=0.52\textwidth]{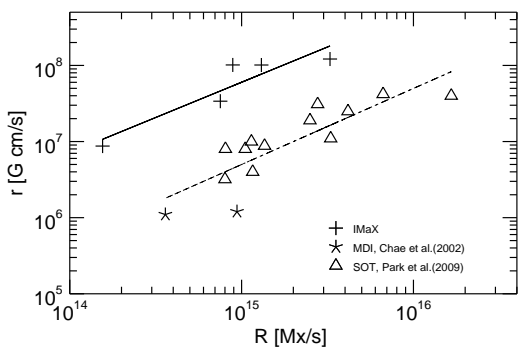}
  \caption{Flux decay rates (R) and specific cancellation rates (r) of Class II cancelling pairs (+ symbols). The asterisks, triangles represent r values from \cite{chb} and \cite{pa}, respectively. The solid line is a regression to the IMaX data, and the dashed line fits data from \cite{pa}. These fits have been plotted mainly to guide the eye.}
\label{fig:g}
\end{figure}

Variation of specific flux cancellation rate with flux decay rate is shown in Fig. \ref{fig:g}. It is evident from the figure that the specific flux cancellation rates of IMaX events are distinctively different from those of the SOT events. The flux decay values are comparable to that  retrieved from SOHO/MDI data \citep{chb, lit, pa}. Of the 12 events studied by \cite{pa} using Hinode/SP data, 6 events have decay rates comparable to ours, while the other 6 have higher decay rates. One thing to note here is that the size of the elements in \cite{pa} is greater than one arcsec as compared with sub-arcsec features in this study. Consequently, the values of the cancellation interface length, in this work, are lower than those obtained by, e.g., \cite{chb}, \cite{pa}. The specific cancellation rate of the events studied here ranges from 9 $\times$ 10$^6~$G cm s$^{-1}$ to 12.1 $\times$ 10$^7~$G cm s$^{-1}$. The mean value of 7.3 $\times$ 10$^7$ G cm s$^{-1}$ is an order of magnitude greater than in \cite{chb}, and about a factor of four greater than in \cite{pa}, for the data from Hinode/SOT. According to \cite{pa}, higher value of $r$ implies that photospheric magnetic reconnection involves either stronger magnetic fields or faster converging motions. We obtained converging speeds that are comparable to the speed range of  granular convection. It could be possible that granular motion is driving magnetic reconnection associated with the Class II pairs. This is different from the existing notion of supergranular motion driving cancellation related reconnection \citep[e.g.,][]{dere}.

\begin{table*}
\caption{Flux decay rates (R), cancellation interface length (l), specific flux cancellation rates (r), and ($B_\mathrm{LOS, max} \times v_\mathrm{conv}$) values for Class II pairs}
\label{tab:t3}
\centering   
\begin{tabular}{c c c c c}
\hline\hline  
\\
 ROI & $R$ & $l$ & $r$ & $B_\mathrm{LOS, max} \times v_\mathrm{conv}$\\
   &[10$^{15}$~Mx s$^{-1}$] & [Mm] & [10$^7$~G cm s$^{-1}$] & [10$^7$~G cm s$^{-1}$]\\
\\
   \hline    
   \\
1  & 0.89  & 0.08  & 10.1 & 5.9\\
2  & 3.3  & 0.27 & 12.1 & 9.8\\
3  & 0.75  & 0.22 & 3.4 & 1.5\\ 
    & 0.15   &0.17 &0.9 & 0.56\\ 
9  &1.3  & 0.13 & 10.1 & 4.56\\
\hline    
\end{tabular}
\end{table*}

Both up- and downflow are found to coexist within the individual magnetic elements of the pairs during some part of the pair's lifetime, a situation similar to that of Class I sample. However, calculation of the average Doppler velocity of each magnetic feature returns downflow values. In short, the pairs display downflows throughout their lifetime. This result is similar to that reported by \cite{kbb} in their study on granular-scale cancellations. Along the PIL between the feature pairs, we observe (1) significant LP signal, and (2) a significant increases in the line core intensity in the beginning followed by a decrease for three of the representative elements. Significant Stokes $V$ continuum signal during cancellation is seen in only a minority of features (two features belonging to two different pairs).

\begin{figure}
\centering
  \includegraphics[trim=190 280 80 235, clip, width=0.52\textwidth]{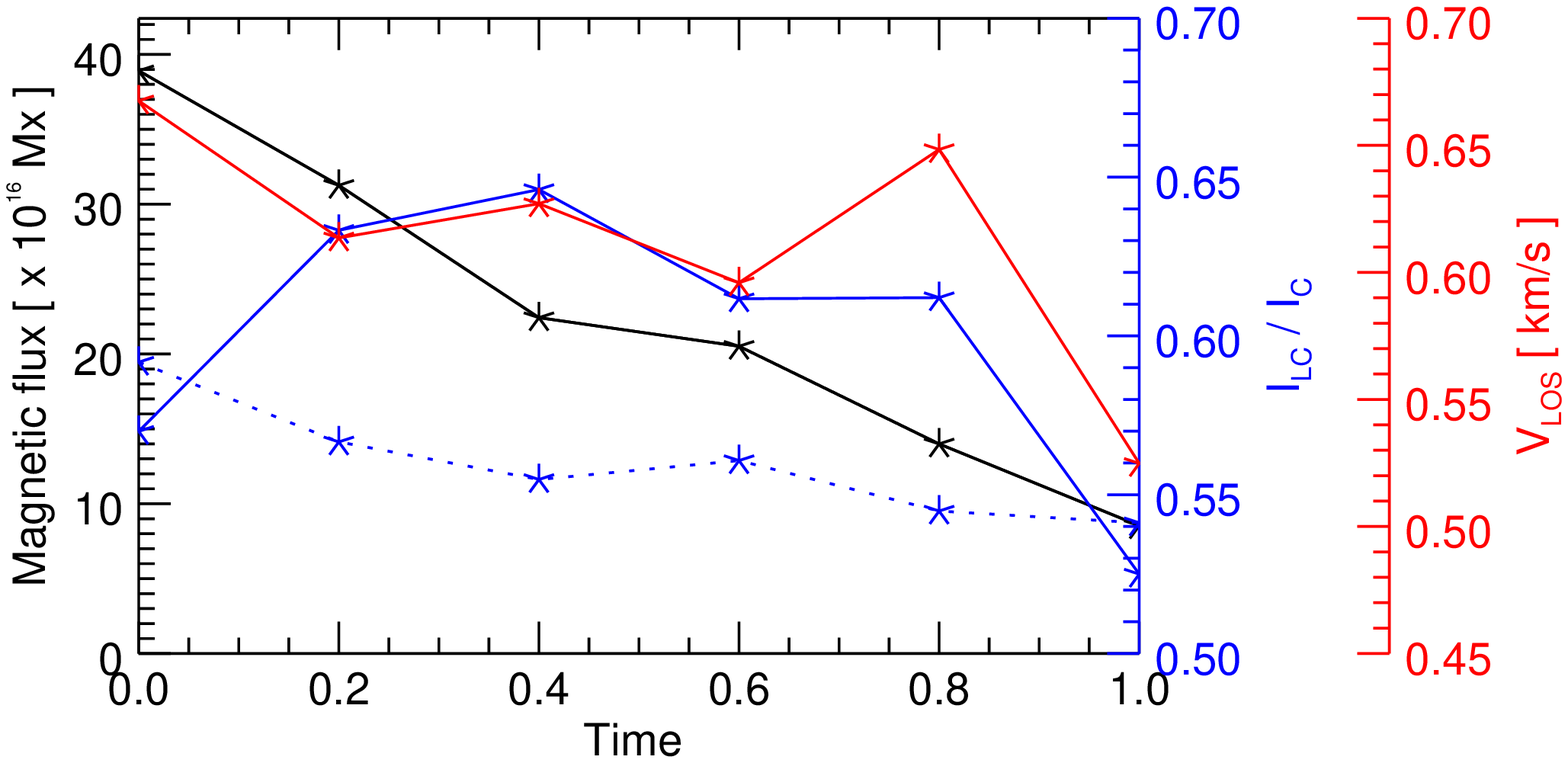}
  \caption{Same as Fig. \ref{fig:d}, but for the 5 cancelling pairs belonging to Class II. Average Doppler velocity of the representative feature (see text for details) is plotted in red. The blue line represents the mean
line core intensity along the PIL.}
\label{fig:h}
\end{figure}

The average magnetic flux evolution of the representative features of the 5 pairs is plotted in Fig. \ref{fig:h} vs. time, normalized to the cancellation lifetimes. As expected, the total magnetic flux decreases over time, with the patches losing roughly $2/3$ of their total magnetic flux in the course of the cancellation. The mean Doppler velocity of the patches (red line in the same figure) is always redshifted and does not change significantly over time. The line core intensity averaged over the PIL for three of the representative features is plotted in blue solid line, which shows an increase in the beginning followed by a decrease. The dashed blue line depicts the same for the remaining two representative features, which demonstrate a monotonous decline of line core intensity.

\section{Summary and Discussion} \label{sec:conc} 
We investigated 11 cancellation events occurring at small scales. The flux decay rate of the 11 events ranges from $-3.3\times 10^{15}$~Mx s$^{-1}$ to $-0.24\times 10^{15}$~Mx s$^{-1}$. The peak flux values of the individual features studied here are one to two order-of-magnitude smaller than in \cite{chb}, and three orders-of- magnitude smaller than in \cite{kba}. Also the cancellation times are less than 10 minutes, as compared with hours reported in, e.g., \cite{chb, chc}. As far as we are aware this is the first attempt to statistically estimate physical parameters of sub-arcsec size cancellation events. We classified the events into two classes, I and II according to whether they take place between pre-existing flux and a newly appeared opposite polarity feature, or happen between previously unconnected features that converge toward each other and cancel. In total we studied 6 Class I and 5 Class II events. The characteristics of the two classes are summarized and discussed as follows:

\underline{Class I}\\
1. Cancellation happens between a small patch and a much bigger opposite polarity patch. The smaller patch is of sub-arcsec size and disappears completely by the end of the cancellation.\\
2. The smaller features have apparent peak LOS field below 300~G and a strongly inclined field throughout their lifetime. We stress, however, that the field may be unresolved and may in reality be higher. Similarly, the LOS inclination is probably overestimated due to noise.\\
3. The total magnetic flux of the small patches peaks at around $10^{17}$~Mx. Their flux decay rate is about $10^{14}$~Mx s$^{-1}$.\\
4. The mean Doppler velocity of the disappearing patch switches from blueshift in the emerging phase to redshift in the decay phase. It should also be noted that the patches enclose both upflow and downflow within them during most of their lifetime. Similarly, along the PIL between the cancelling opposite-polarity pairs Doppler velocities in both directions are found.\\
5. An enhanced LP signal is observed along the PIL for at least some time while cancellation is proceeding.\\
6. During cancellation, we observe an increase in the line core intensity along the PIL for all the events. This increase corresponds to a temperature enhancement of $60 - 230$~K in the middle photosphere.\\
7. All these events except the ones in ROI 2 and 5 are covered by the SuFI FOV. No visible increase in SuFI intensity images is observed. In particular, no significant enhancement in \ion{Ca}{ii} {H} core intensity is seen. This implies that the temperature enhancement present in the middle photosphere does not extend into the lower chromosphere.

\begin{figure*}
\centering
   $\begin{array}{rl}
  {\includegraphics[width=0.8\textwidth]{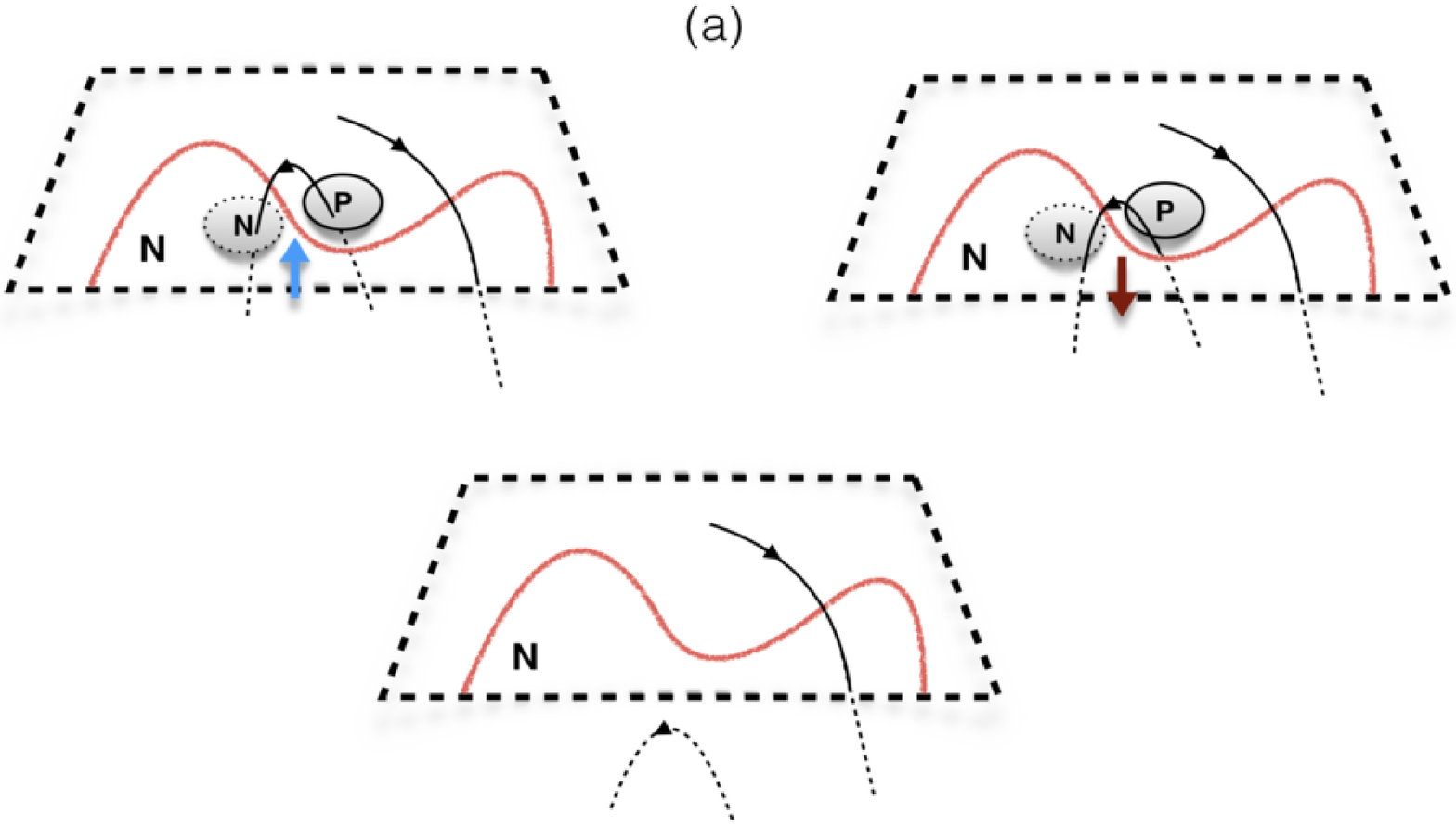}}\\
   {\includegraphics[width=0.8\textwidth]{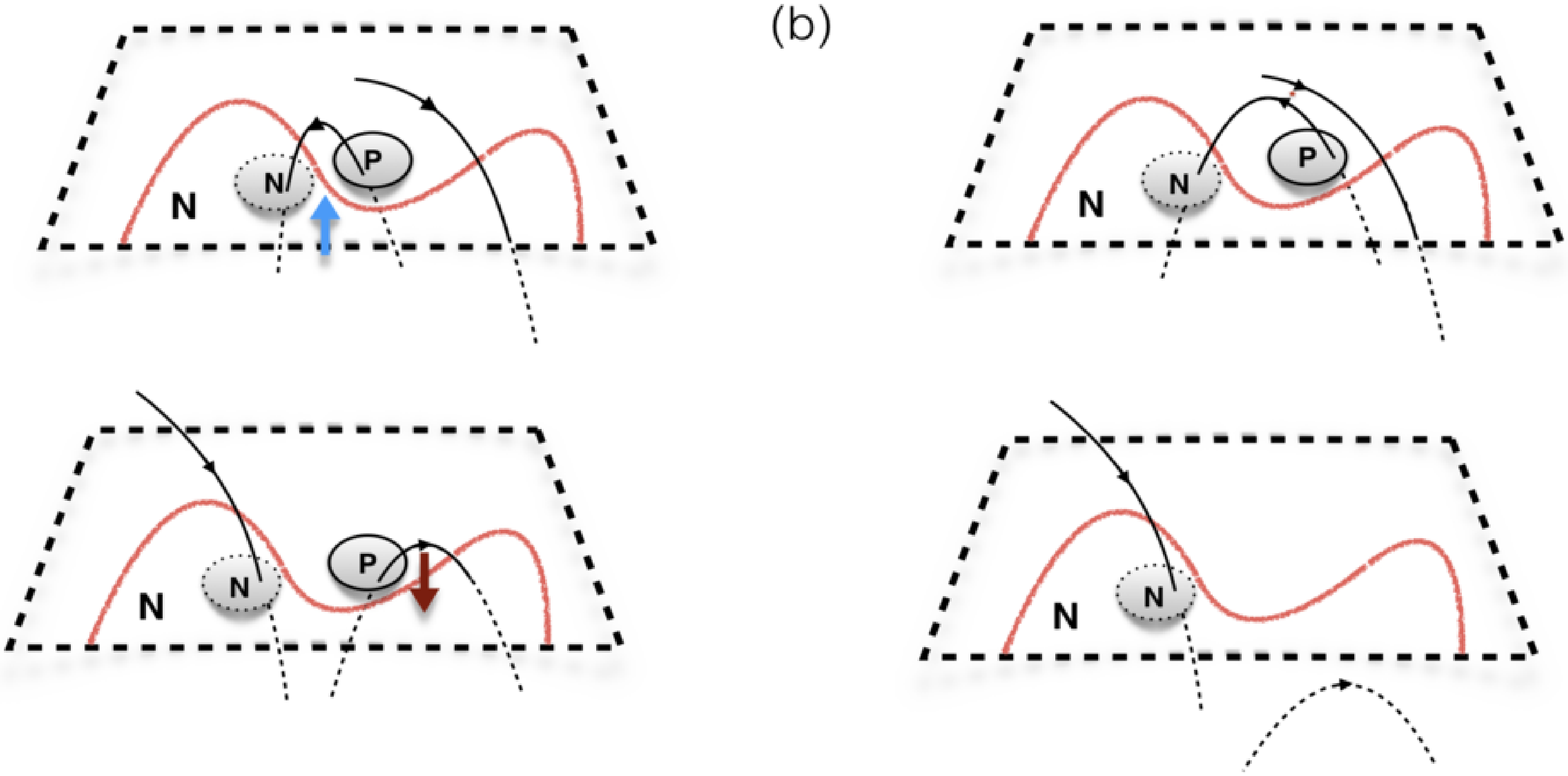}}
\end{array}$
  \caption{Cartoon of the possible scenarios underlying Class I cancellation events. (a) Emergence followed by flux retraction. (b) Emergence, then reconnection followed by submergence. The bold dashed rectangle represents a part of the solar surface. N and P indicate negative and positive magnetic polarity. The solid and dotted ovals represent the newly emerged bipole. The red line represents the boundary of the pre-existing large magnetic feature. Solid black lines are parts of field lines above the solar surface, dashed lines are parts below it. Blue and red arrows indicate up- and downflows.}
  \label{fig:i}
\end{figure*}

We propose two scenarios that could result in Class I events: (1) submergence events (see Fig. \ref{fig:i} a) in which a pair of magnetic features rises to the surface very close to a pre-existing bigger magnetic feature. Hence the newly appearing patch with the same polarity as the bigger feature is masked by the latter. The visible member of the pair glides along with the convection pattern and is finally dragged down below the surface by the downdraft along with the other, hidden member. In this scenario, the apparent rise in line core intensity and temperature along the PIL is produced by the larger patch which has strong magnetic fields with rather vertical orientation. The apparent brightness enhancement at the PIL could be nothing more than the brightening due to the large magnetic feature, which expands with height. Also, the change in sign of the Doppler velocity could be related to the cancellation, but may have more to do with the motion of the features from granules, where they emerge \citep{lt, dan}, to intergranules, where magnetic features spend most of their life. \cite{lim} reported a similar scenario in which cancellation and subsequent disappearance of sub-granular scale magnetic elements was observed near an active region. The authors proposed that the cancellation could be due to either the emergence of U-loops, or submergence of $\Omega$-loops, and may not be due to magnetic reconnection as they didn't observe any chromospheric signal associated with the event. 

The second scenario involves reconnection followed by submergence (see Fig. \ref{fig:i} b). In this scenario things start the same way as in scenario 1, but instead of a simple retraction of the bipole, a part of the field of the large feature reconnects with the freshly emerged opposite polarity feature prior to its submergence. To check whether the observed rise in the line core intensity along the PIL can be produced by reconnection, we made a simple estimation for the pair in region 7: (a) The maximum magnetic energy available through flux cancellation is roughly estimated to be, $\delta E_\mathrm{m} =  2 (B_\mathrm{max}^2/8\pi) Area_\mathrm{max}h$, where $h$ is the height range over which the heating takes place. $B_\mathrm{max}$ and $Area_\mathrm{max}$ are taken from Table \ref{tab:t1} after accounting for the inclination of the field. The main uncertainty lies in the determination of $B_\mathrm{LOS}$ and $\gamma_\mathrm{LOS}$. Using the values given in Table \ref{tab:t1}, we get $B = 540$~G. The factor 2 accounts for the total energy available from both polarities, assuming that the flux lost from the bigger feature has the same field strength as the smaller feature. Assuming $h$ to be about 200~km (very roughly the height range over which the line core forms at the spectral resolution of the IMaX magnetograph), we get $\delta E_\mathrm{m} = 2.0 \times 10^{26}$ erg. Typically, only a fraction of this energy is available as free energy that can be released via reconnection. If we assume that it is 10\% of the total magnetic energy, a not unusual value, then we have $E_\mathrm{free} = 2.0 \times 10^{25}$ erg available. (b) The thermal energy needed for the heating can be very roughly approximated as $\delta E_\mathrm{th} =  3/2nk_\mathrm{B}\delta TAh$, where $n$ is the number density at $log\tau = \mathrm{-2.5}$, $k_\mathrm{B}$ is the Boltzmann constant, and $A$ is the area of the PIL over which the heating takes place (averaged over the duration for which the intensity enhancement is visible). $\delta T$ is the change in temperature along the PIL at $log\tau = \mathrm{-2.5}$ during the period of cancellation. With $n \sim 10^{16}$~cm$^{-3}$, $\delta T \sim 80$~K and $A \sim 3.0 \times 10^{14}$~cm$^{2}$, we get $\delta E_\mathrm{th} \sim 10^{24}$~erg; (c) since the enhanced temperature is maintained for some time, we also need to consider the energy radiated over the time $\delta t$ over which the temperature (and line core brightness) is enhanced, $\delta E_\mathrm{r} =  4\sigma T^3\delta TA\delta t$, where $\sigma$ is the Stefan-Boltzmann constant, $T$ is the mean quiet-sun temperature at $log\tau = \mathrm{-2.5}$. Taking $T$ to be about 4500~K and $\delta t \sim 73$~s (see Fig. \ref{fig:c}), we get $\delta E_\mathrm{r} = 3.6\times10^{25}$~erg.

The magnetic free energy available from cancellation is an order of magnitude greater than the thermal energy enhancement. In addition, $\delta E_\mathrm{r}$ is comparable to $\delta E_\mathrm{free}$, so that the latter can account for the excess energy radiated during cancellation. This energy estimation is consistent with a reconnection scenario in which basically all the magnetic free energy in the reconnecting field is released prior to submergence. The line core intensity and temperature enhancement along the PIL associated with all the Class I events could be the result of magnetic reconnection. If that is the case, these reconnection events would be happening close to the photosphere, since we do not see any significant brightness enhancement in the SuFI 397 channel which corresponds to the lower chromosphere. 

Even more energy can be released if Ohmic heating along the current sheet between the pre-existing large magnetic structure and the newly emerged small opposite-polarity element, directly converts magnetic energy into heat. Also in this case, the heating found in the middle photosphere could be explained by conversion of magnetic energy into heat. 

If we have strongly overestimated the field strength, then only the presence of Ohmic heating at a current sheet can explain the heightened temperature, unless the free energy constitutes a much larger fraction of the total magnetic energy. One hint that the field strength may indeed be overestimated is the large magnetic inclinations. Due to buoyancy, kG  magnetic fields tend to be close to vertical in the photosphere. The large inclinations, if real, tend to support a weaker field. If the large inclinations are overestimates, then, however, the ratio of the field strength to its LOS component is smaller than estimated above.

\underline{Class II}\\
1. All the pairs except the ones in region 1 and 2 are of sub-arcsec size. The features involved lose a large fraction of their flux (34\% - 97\% of their peak flux), but are generally still detectable at the end of the cancellation.\\
2. Peak LOS fields are below 300~G and the fields of all the elements are highly inclined, except for the ones in regions 1 and 2. The same caveats apply as for the magnetic vector determinations for Class I events.\\
3. The total magnetic flux peaks between 10$^{17}$~Mx and 10$^{18}$~Mx. The flux decay rate is about  10$^{15}$~Mx s$^{-1}$.\\
4. Similar to the Class I events, here also the individual cancelling magnetic features enclose both upflows and downflows over most of their lifetime. Also, along the PIL both redshift and blueshift are observed. However, on average the downflows tend to dominate within the cancelling magnetic patches.\\
5. All 5 events show significant LP signal along the PIL for at least some time during the event.\\
6. Enhanced line core intensity along the PIL was observed for 3 out of the 5 events.\\
7. Since no SuFI data were available for these features, we cannot say whether the low chromosphere was heated or not during these cancellation events.

We propose that magnetic reconnection associated with Class II events is probably driven by granular motions as opposed to supergranulation, which, in earlier studies, was proposed to drive large-scale cancellations. Specific flux cancellation rates of the Class II events seem to indicate that they belong to a somewhat different category of cancellations when compared with those studied in SOT/Hinode and MDI/SOHO data. Information of magnetic field-line connectivity of Class II events is absent due to the lack of magnetic field data sampling higher layers of the atmosphere. And hence we are unable to say at what height the reconnection is happening. This is a question we would like to address in the future using data from different atmospheric layers. Since, e.g., in the event in ROI 9, the line core brightening starts well before the two features meet (as seen in the lower photosphere), we speculate that reconnection already starts when the field, which expands rapidly with height, meets in the upper photosphere. Such a scenario has been simulated by \cite{cam}. We hope to be able to sample the necessary height-range with the instrumentation onboard the planned third flight of Sunrise.

The Doppler velocity within the features or along the PIL during cancellation is not unique to that region, rather it follows velocity pattern of surrounding convective motions. And, the mean velocity values fall within the range of typical granular-scale convective motions. This could be an indication that the derived velocity values represent the motion of the nearby gas undergoing convection, rather than that along the field lines of features undergoing cancellation, as pointed out by \cite{kbb}. We propose that the evolution of these events is guided by the nearby granular flows. In short, it is difficult to disentangle the contribution to Doppler velocity of cancellation from convective flows.

We observed substantial LP signal along the PIL in all the 11 samples, whereas \cite{kbb} reported the presence of LP signal in only one of the 5 events they studied. The authors attributed the absence of LP signal to the limited spatial resolution of Hinode SOT/SP. Their dataset had a cadence of 5.5 minutes and a spatial resolution of 0.32\arcsec. Ours has a cadence of 36.5 sec and the LP maps are obtained after a spatial binning of 3 $\times$ 3 pixels, which means the spatial resolution of our LP signal is similar to that of \cite{kbb}. However, the enhanced LP signal was seen for less than 5 minutes for all the events reported in this study, so that temporal cadence is likely one of the deciding factors in detecting the presence of LP signals during cancellation. 

A combination of many-line spectropolarimetric (to seamlessly cover a broader height range, but also to obtain a better S/N ratio) and high cadence magnetographic data are expected to provide more information. Also, numerical simulations that can mimic such small-scale events will help to understand the dynamics involved in their emergence, evolution and final disappearance.

\begin{acknowledgements}
A.J.K thanks, R. Cameron and L.S. Anusha for helpful suggestions and comments. We also thank an anonymous referee for comments that significantly improved the manuscript. The German contribution to \sunrise{} and its reflight was funded by the Max Planck Foundation, the Strategic Innovations Fund of the President of the Max Planck Society (MPG), DLR, and private donations by supporting members of the Max Planck Society, which is gratefully acknowledged. The Spanish contribution was funded by the Ministerio de Econom\'{i}a y Competitividad under Projects ESP2013-47349-C6 and ESP2014-56169-C6, partially using European FEDER funds. The HAO contribution was partly funded through NASA grant number NNX13AE95G. This work was partly supported by the BK21 plus program through the National Research Foundation (NRF) funded by the Ministry of Education of Korea. 
\end{acknowledgements}

\bibliography{bib}

\end{document}